\documentclass[3p,times,procedia]{elsarticle}
\usepackage{nupha_ecrc}

\volume{00}

\firstpage{1}

\journalname{Nuclear Physics A}

\runauth{A. Rustamov}

\jid{nupha}

\jnltitlelogo{Nuclear Physics A}

\usepackage{amssymb}

\usepackage[figuresright]{rotating}
\usepackage{mathtools}
\newcommand{\eq}[1]{\begin{align} #1 \end{align}}
 
\newcommand{\Prom}{\mathrm{p}}

\begin{document}

\begin{frontmatter}



\dochead{XXVIIIth International Conference on Ultrarelativistic Nucleus-Nucleus Collisions\\ (Quark Matter 2019)}

\title{ Overview of fluctuation and correlation measurements}


\author[GSI,NNRC]{A.~Rustamov}
\ead{a.rustamov@cern.ch}
\address[GSI]{GSI Helmholtzzentrum f{\"u}r Schwerionenforschung, Darmstadt, Germany}
\address[NNRC]{National Nuclear Research Center, Baku, Azerbaijan}

\begin{abstract}
One of the ultimate goals of nuclear collision experiments at high energy is to map the phase diagram of strongly interacting matter. A very challenging task is the determination of the QCD phase structure including the search for critical behavior and verification of the possible existence of a critical end point of a first order phase transition line. A promising tool to probe the presence of critical behavior is the study of fluctuations and correlations of conserved charges since, in a thermal system, these fluctuations are directly related to the equation of state (EoS) of the system under the study. In this report an overview is given of several experimental measurements on net-proton multiplicity distributions such as cumulants and multi-particle correlation functions. 

\end{abstract}

\begin{keyword}
Quark-gluon plasma \sep Fluctuations and correlations \sep Conservation laws


\end{keyword}

\end{frontmatter}


\section{Introduction}
\label{sec:intro}
The study of a phase structure of strongly interacting matter is the focus of many research activities worldwide. As the theory of strong interactions, Quantum Chromodynamics (QCD), is asymptotically free, in the realm of high temperature and/or density the fundamental degrees of freedom of the strong interactions come into play. By colliding heavy-ions at different energies one hopes to heat and/or compress the matter to energy densities at which a transition from matter consisting of confined baryons and mesons to a state of liberated quarks and gluons (deconfined phase) begins. However, liberated quarks and gluons are not what one ultimately observes in experiments. The subsequent expansion and cooling of the deconfined phase leads to formations of  hadrons, which fly outwards, and get registered by the detectors.  This process of hadronization plays a key role in understanding what detectors see. The headway is to establish a bridge between the events which occur before the hadronization and the experimental outcome. 
The situation is much similar to reconstruction of the cosmological Big Bang from observables like Hubble expansion, the cosmic microwave background and the abundance of light atomic nuclei.

Phase transitions are usually studied by looking to the response of the system to external perturbations. For example, the liquid gas phase transition can be probed by the response of the volume to a change in pressure, which is encoded in the isothermal compressibility. In the Grand Canonical Ensemble (GCE) formulation of statistical mechanics the latter  contains fluctuations of liquid constituents from microstate to microstate. Hence, the objective is to relate macroscopic parameters of the system, which define its EoS, with its microscopic details encoded in fluctuations.

In a similar way, phase transitions in strongly interacting matter can be addressed by investigating the response of the system to external perturbations via measurements of fluctuations of conserved charges such as baryon number or electric charge~\cite{Koch:2008ia, Adamczyk:2013dal}. 

For a thermal system of volume $V$ and temperature $T$, within
the Grand Canonical Ensemble, fluctuations of a given net-charge
$\Delta N_{B}=N_{B}-N_{\bar{B}}$ are related to the corresponding
reduced susceptibilities $\hat{\chi}_{n}^{B}$~\cite{Bazavov:2020bjn}:

\begin{equation}
\frac{1}{VT^{3}} \kappa_{n}(\Delta N_{B})  = \hat{\chi}_{n}^{B},
\label{cumulants}
\end{equation}
with $\hat{\chi}_{n}^{B}$ defined as $n^{th}$  derivative of the
reduced thermodynamic pressure $\hat{p}\equiv \frac{p}{T^{4}}$ with
respect to the corresponding reduced chemical potential
$\hat{\mu}_{B}\equiv\frac{\mu_{B}}{T}$ and $k_{n}(\Delta N_{B})$ stands for cumulants of conserved net-charge distributions. 

At LHC energies there would be, for vanishing light quark masses (u and d quarks),  a temperature-driven  second order phase transition between a hadron gas and a quark--gluon plasma~\cite{Stephanov:2004wx}. For realistic quark masses this transition becomes a smooth cross over~\cite{LQCD1,LQCD2}. Nevertheless, because of the small masses of the light current quarks,  one can still probe critical phenomena at LHC energies (vanishing baryon chemical potential) as reported in~\cite{Friman:2011pf}. Indeed, recent LQCD calculations \cite{LQCD1,LQCD2} exhibit a rather strong signal for the existence of a pseudo-critical chiral temperature of 156.5$\pm$1.5 at $\mu_{B}$ = 0.  Moreover, this pseudo-critical temperature turns out to be in good agreement with the chemical freeze-out  temperature as extracted by the analysis of hadron multiplicities measured by the ALICE experiment~\cite{Andronic:2017pug,Andronic:2018qqt}. This implies that the strongly interacting matter created in central collisions of Pb nuclei at LHC energies freezes out  near the chiral phase transition line. Within statistical uncertainties, the pseudo-critical line is also consistent with freeze-out temperatures determined by the STAR BES-I data~\cite{Adamczyk:2017iwn}. 

At larger values of $\mu_B$ it is generally expected that a line of first order phase transition exists, which ends in a second order chiral critical point (cf.~\cite{Stephanov:2004wx} and references therein).

To increase  sensitivity, it is better to exploit higher order cumulants because they are better messengers of long range correlations and large fluctuations in the proximity of the critical point. 

Two comments are in order here. First, the LQCD calculations are predicted for a thermal system in a fixed volume. While the notion of the volume is not firmly defined in the experimental case, it is a common practice  to use number of wounded nucleons as a proxy for the reaction volume (within the wounded nucleon model). In experiments, however, wounded nucleons always fluctuate from event to event, hence direct comparison with LQCD becomes challenging. Second, as mentioned above, the LQCD calculations are performed within the GCE formulation of the statistical mechanics, where net-baryons are not conserved in each microstate. However, in experiments baryon number is conserved in each event. While for the analysis of mean multiplicities the appropriate acceptance can be selected in order to fulfill the requirements of the GCE, for the higher moments there does not exist any a priory prescription for selecting the "required" acceptance. Indeed, if the selected acceptance window is too small, possible dynamical correlations will be washed out and net-baryons will be distributed according to the Skellam distribution originating from Poisson distributions for single baryons due to small number statistics. For a larger acceptance, however, subtle contributions to measured second cumulants can become observable as for instance those coming from baryon number conservation. Hence the acceptance has to be selected large enough to avoid the small number Poissonian limit and final results should be corrected for contributions originating from participant fluctuations and conservation laws.

Interestingly, contributions from participant fluctuations to the second and third order cumulants of net-baryon distributions are found to vanish at mid-rapidity for LHC energies while higher cumulants of even order are non-zero even when the net-baryon number at mid-rapidity is zero~\cite{Braun-Munzinger:2016yjz, Skokov:2012ds, Acharya:2019izy}. 

\section{Basic notations}
The $r^{th}$ central moment of a discrete
random variable $X$, with its probability distribution $P(X)$, is
generally defined as

\begin{equation}
\mu_{r} \equiv  \left<\left(X-\left<X\right>\right)^{r}\right> = \sum_{X}\left(X-\left<X\right>\right)^{r}P(X),\\
\label{cum_definition1}
\end{equation}

where $\left<X\right>$ denotes the mean of the distribution

\begin{equation}
\left<X\right> =  \sum_{X}XP(X) .\\
\label{mean_definition1}
\end{equation}

In a similar way we introduce moments about the origin, thereafter referred to as raw moments 

\begin{equation}
 \left<X^{r}\right> = \sum_{X}X^{r}P(X).\\
\label{mean_definition2}
\end{equation}

The cumulants of $X$ are defined as the coefficients in the Maclaurin
series of the logarithm of the characteristic function of $X$.  The
first four cumulants read
 \eq{\label{kumulants_definition}
  &\kappa_{1} = \left<X\right> , \nonumber\\ &\kappa_{2} = \mu_{2} =
  \left<X^{2}\right> - \left<X\right>^{2} , \nonumber\\ &\kappa_{3} =
  \mu_{3} = \left<X^{3}\right> - 3\left<X^{2}\right>\left<X\right> +
  2\left<X\right>^{3}, \\ &\kappa_{4} = \mu_{4} - 3\mu_{2}^{2} =
  \left<X^{4}\right> - 4\left<X^{3}\right>\left<X\right> -
  3\left<X^{2}\right>^{2} \nonumber\\ &
  +12\left<X^{2}\right>\left<X\right>^{2} -
  6\left<X\right>^{4}. \nonumber }




In the following, the $X$ quantity is replaced by the net-proton number ($n_{p}-n_{\bar{p}}$) which is used as a proxy for net-baryons.

\begin{figure}[htb]
\centering
\includegraphics[width=1.\linewidth,clip=true]{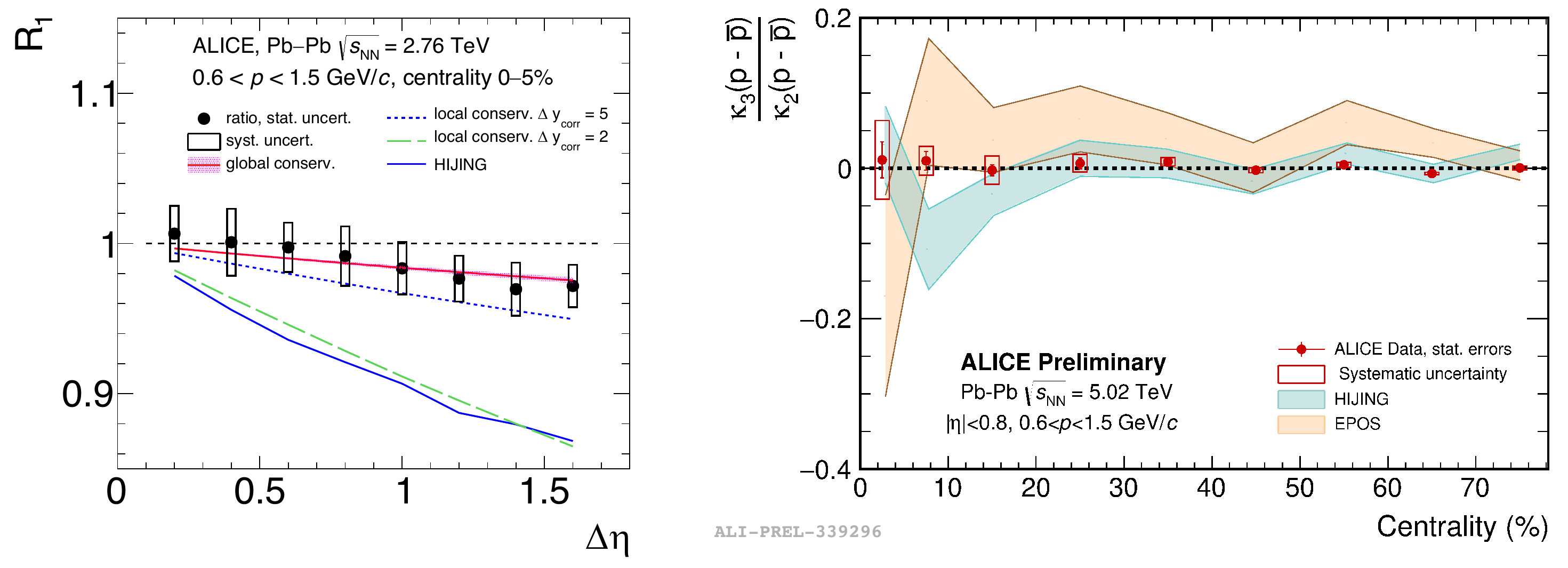} 
\caption{ALICE Pb--Pb data. \emph{Left panel}: Pseudorapidity  dependence of the normalized second cumulants of net-protons R$_1$ at $\sqrt{s_{\mathrm {NN}}} = 2.76$~TeV. Global baryon number conservation is depicted as the pink band. The dashed lines represent the predictions from the model with local baryon number conservation~\cite{Braun-Munzinger:2019yxj}. The blue solid line, represents the prediction using the HIJING generator. \emph{Right panel}: Centrality  dependence of the ratio of third to second order cumulants for net-protons at $\sqrt{s_{\mathrm {NN}}} = 5.02$~TeV. The ALICE data are shown by red markers, while the colored shaded areas indicate the HIJING and EPOS model calculations.
}
\label{fig_ALICE}
\end{figure} 

By their definition, cumulants are extensive quantities, i.e., are proportional to the system volume. To remove the volume dependence normalized cumulants R$_{1}$, $S\sigma$  and $k\sigma^{2}$ are introduced

\begin{equation}
\mathrm{R}_{1}= \kappa_{2}(n_{\Prom} - n_{\overline{\Prom}})/\left< n_{\Prom} + n_{\overline{\Prom}} \right>,\,\,\ 
S\sigma= \kappa_{3}/\kappa_{2},\,\,\ k\sigma^{2}=\kappa_{4}/\kappa_{2},
\label{normCum}
\end{equation}
where $S$ and $k$ denote the skewness and kurtosis of the distribution.

In general all cumulants also depend on volume fluctuations. Before taking the ratios introduced in Eq.~\ref{normCum} the contributions from volume fluctuations have to be accounted for. Only in this case these ratios depend neither on  volume no on its fluctuations. In particular, at lower beam energies it is essential to remove contributions from volume fluctuations~\cite{Braun-Munzinger:2016yjz,  Adamczewski-Musch:2020slf}.

\section{Experimental results}
\subsection{Results from ALICE}

In the left panel of Fig.~\ref{fig_ALICE} the acceptance dependence of the efficiency corrected normalized cumulants $R1$ (cf. Eq.~\ref{normCum}), measured in Pb--Pb collisions at $\sqrt{s_{\mathrm {NN}}} = 2.76$~TeV,  are presented~\cite{Rustamov:2017lio, Acharya:2019izy}. As already  mentioned in the introduction, at LHC energies the $R1$ values are not affected by volume and its fluctuations.   The analysis is performed with the Indentity Method~\cite{Gazdzicki:2011xz, Gorenstein:2011hr, Rustamov:2012bx, Arslandok:2018pcu} in eight pseudorapidity regions ranging from $-0.1 <\eta<0.1$ up to $-0.8<\eta<0.8$. The data exhibits linear approach to unity with decreasing acceptance,  consistent with predictions based on the assumption of global baryon number conservation~\cite{Bzdak:2012an, Braun-Munzinger:2016yjz}. When imposing a finite acceptance cut the subtle correlations between baryons and anti-baryons, induced by the global baryon number conservation law, weakens.  In the limit of small acceptance these correlations become not visible anymore in the measured second order cumulants. However, the amount of correlation inside finite acceptance depends also on the correlation length $\Delta y_{corr}$ in the rapidity space. This local baryon number conservation~\cite{Braun-Munzinger:2019yxj} would lead to further suppression of the measured R$_{1}$ values. Close inspection of Fig.~\ref{fig_ALICE}, however, indicates that within experimental uncertainties the ALICE data are best described with the large correlation length in the rapidity space, i.e., the observed correlations, to a large extent, are induced by global baryon number conservation. The latter corresponds to the correlation length of $\Delta y_{corr}=2|y_{beam}|$. The HIJING results~\cite{Gyulassy:1994ew}, on the other hand, underestimate the experimental data and correspond to correlation length of $\Delta y_{corr} = 2$. The large correlation length observed in the data takes place before the time~\cite{Dumitru:2008wn}

\begin{figure}[htb]
\centering
\includegraphics[width=1.\linewidth,clip=true]{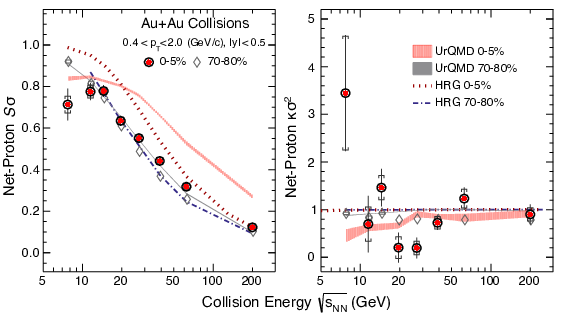} 
\caption{STAR Au--Au data.  The measured $S\sigma$ (\emph{left} panel) and $k\sigma^2$ (\emph{right} panel) values as a function of collision energy for net-proton distributions. The results are shown for 0-5\% and 70-80\% collisions within 0.4 $< pT <$ 2.0 GeV/c and $|y| <$ 0.5. The error bars and caps show statistical and systematic uncertainties, respectively. 
}
\label{fig_STAR}
\end{figure}

\begin{figure}[htb]
\centering
\includegraphics[width=1.\linewidth,clip=true]{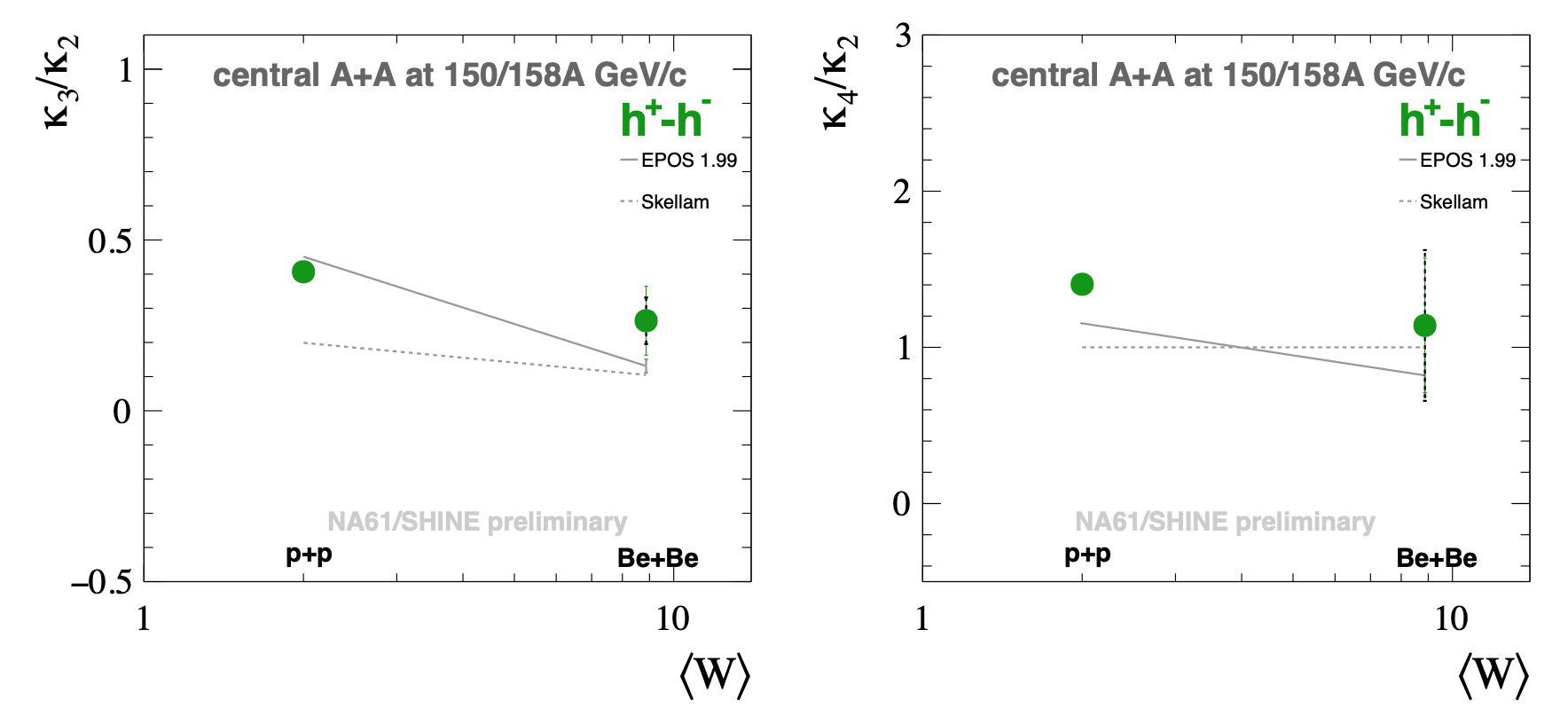} 
\caption{ Preliminary NA61/SHINE results on $\kappa_{3}/\kappa_{2}$ (\emph{left panel}) and $\kappa_{4}/\kappa_{2}$ (\emph{right panel}) ratios on net-charge distribution at beam momenta of 150/158A GeV/c as a function of the mean number of wounded nucleons $\left<W\right>$.}
\label{fig_NA61}
\end{figure} 

\begin{figure}[htb]
\centering
\includegraphics[width=0.4\linewidth,clip=true]{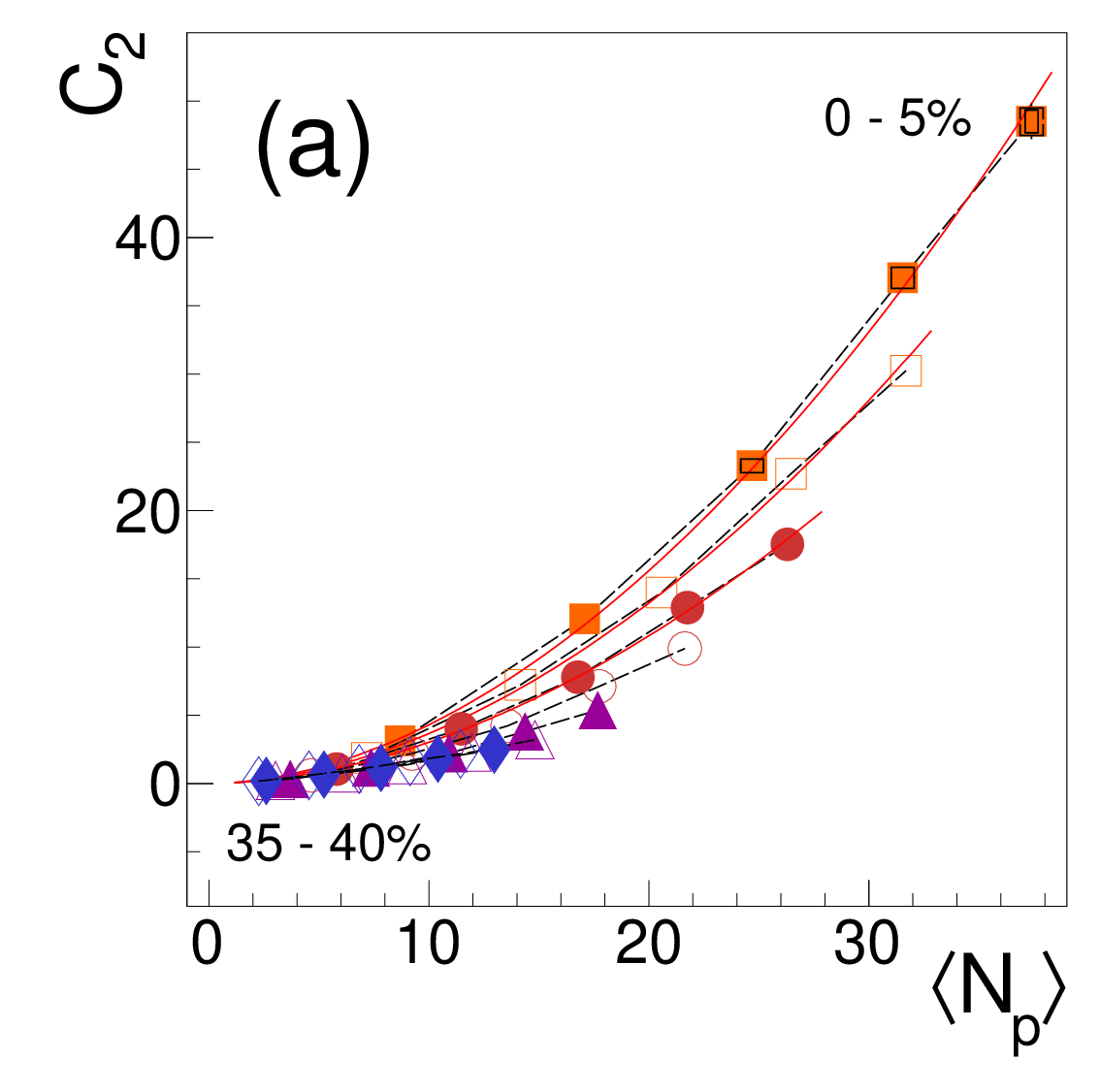} 
\includegraphics[width=0.4\linewidth,clip=true]{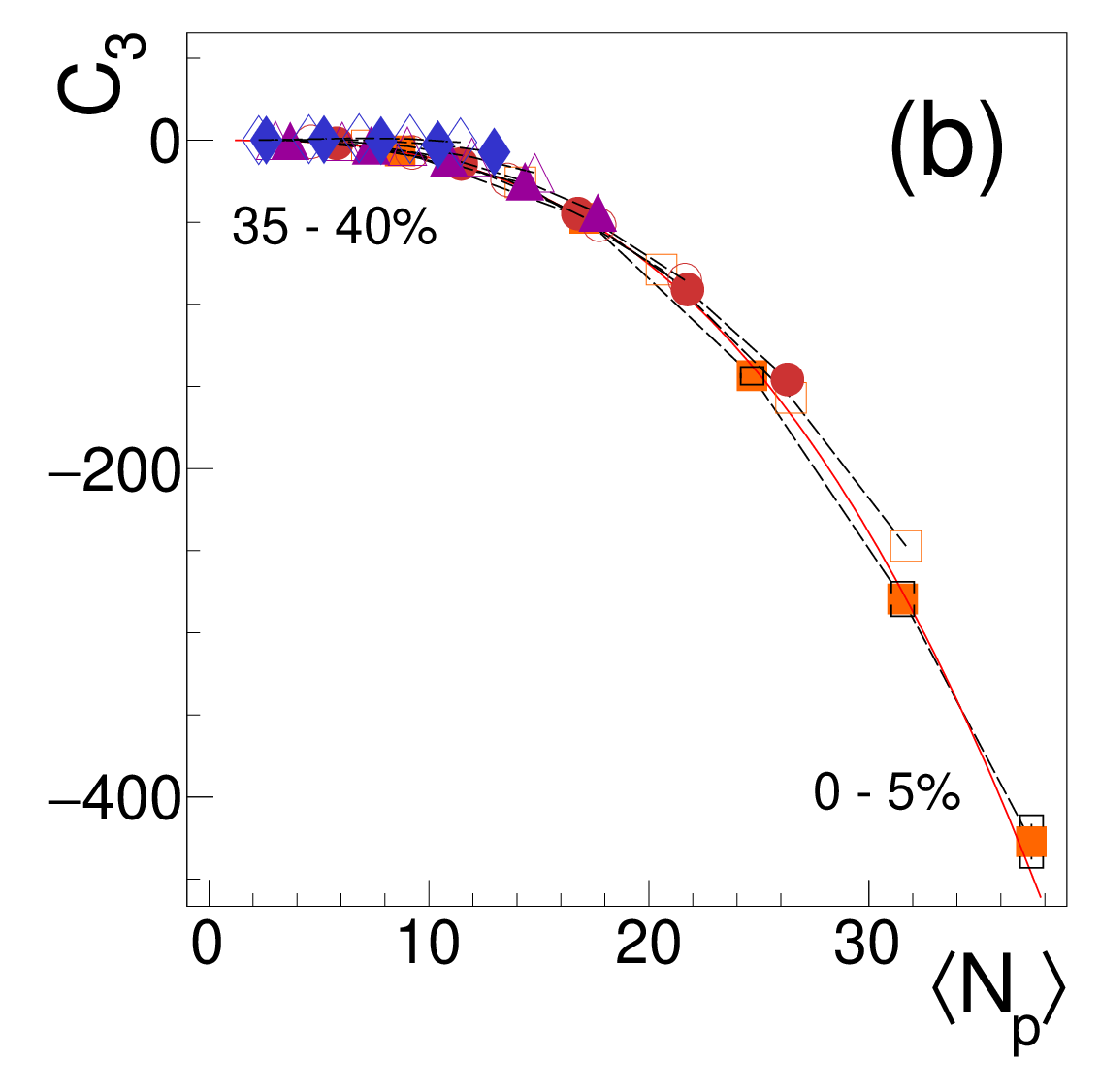} 
\includegraphics[width=0.4\linewidth,clip=true]{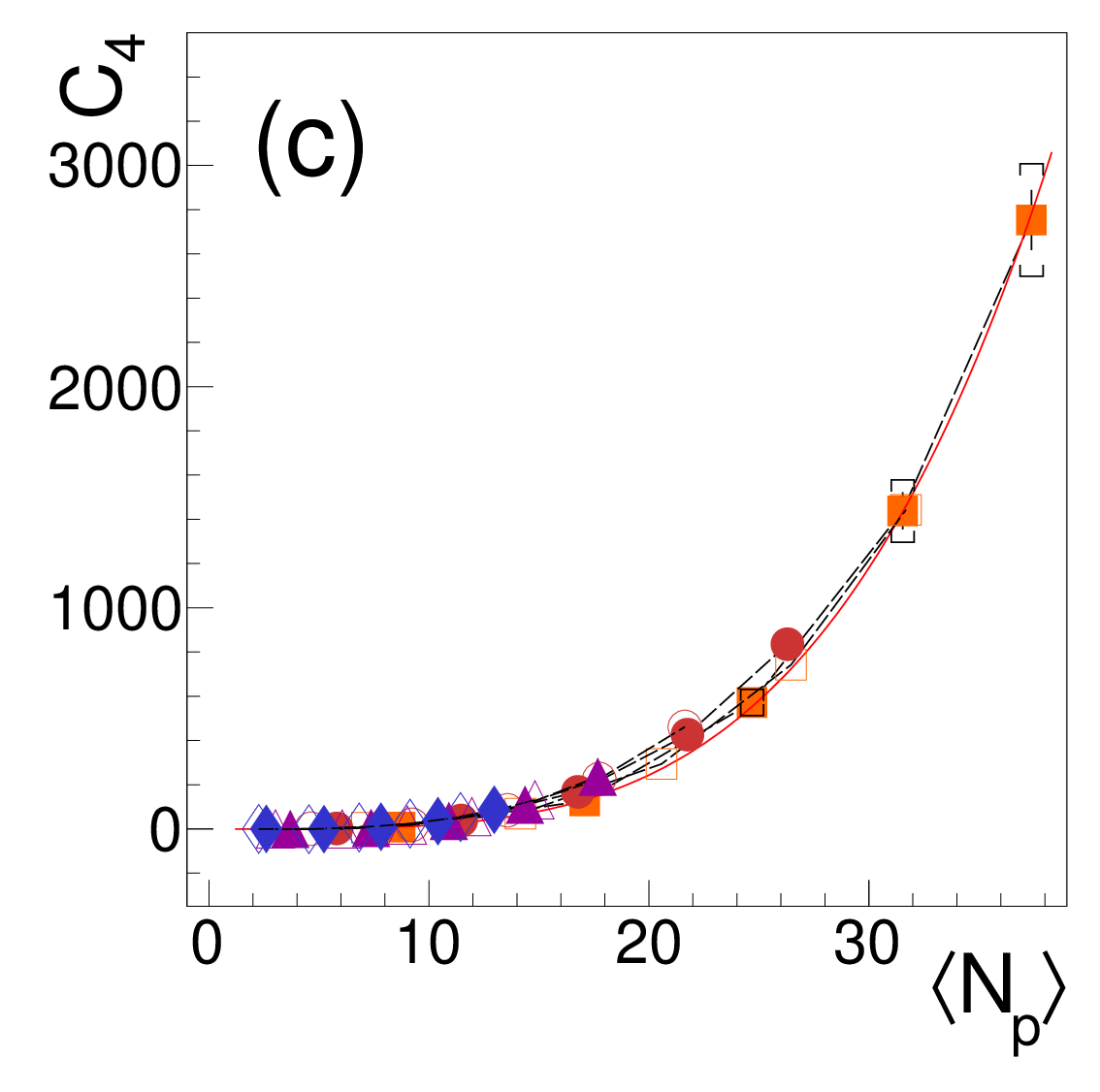} 
\caption{HADES Au--Au data: Efficiency and volume corrected proton correlators $C_{n}$ ($n=2, 3, 4$) as a function of the mean number of protons $\left<N_{p}\right>$ within the selected phase-space bin, $y \in y_{0}\pm \Delta y$ ($\Delta y$ = 0.1,...,0.5) and 0.4 $\leq p_{t} \leq$ 1.6 GeV/c, and for eight centrality selections. Error bars on data are statistical, cups delimit systematic uncertainties (shown, for clarity, only on the 0-5 \% selection). Black dashed lines connect the data points in a given centrality selection and red solid curves are power-law fits $C_{n} \propto \left<N_{p}\right>^{\alpha}$.}
\label{fig_HADES}
\end{figure}

\begin{equation}
\tau_{i}\le\tau_{f}exp(-|\Delta y|/2),
\label{init_time}
\end{equation}
where $\tau_{f}$ denotes the hadron freeze-out time and is of order $\tau_{f}=10 fm/c$ for central Pb--Pb collisions at LHC~\cite{Aamodt:2011mr}. This implies that long range rapidity correlations ($\Delta y_{corr}=2|y_{beam}|$), observed in ALICE data, can only be created at early times, shortly after the collision. 

For symmetry reasons all odd cumulants of net-proton distribution at LHC energies, measured at mid-rapidity, are vanishing. Indeed, the efficiency corrected third cumulants of net-protons as measured by the ALICE collaboration and presented in the right panel of Fig.~\ref{fig_ALICE} are, with the precision below 5\%, consistent with zero~\cite{Arslandok:2020mda}. The consistency with the expected baseline indicates that all  techniques and correction procedures applied to the experimental data are under control and establishes a solid approach to address the higher order cumulants. Indeed, the  critical behavior at the LHC energies is predicted for higher order cumulants of net-baryon distributions~\cite{Almasi:2017bhq}.
A factor of 10 more data, collected in 2018, are already sufficient to address the fourth order cumulans. From 2021 onward a factor of 100 more statistics will be recorded, which allows for precise measurements of six order cumulants. 

\subsection{Results from STAR}

The excitation functions of the efficiency corrected normalized cumulants ($S\sigma=\kappa_{3}/\kappa_{2}$ and $k\sigma^{2}=\kappa_{4}/\kappa_{2}$) in Au--Au collisions as measured by the STAR collaboration~\cite{Adam:2020unf}, are presented in Fig.~\ref{fig_STAR}. The measurements are performed inside the sub-rage of the space space by imposing selection criteria on transverse momentum and rapidity of protons and anti-protons $0.4 < p_{T}(GeV/c) <2.0$ and $|y|<0.5$. The values of $\kappa_{3}/\kappa_{2}$ (cf. left panel of Fig.~\ref{fig_STAR}) are systematically below the ideal HRG baseline in the Boltzmann limit, both for central (0-5\%) and peripheral (70-80\%) collisions. This behaviour is in line with the baryon number conservation effects as reported in~\cite{Braun-Munzinger:2018yru}. On the other hand, the UrQMD~\cite{Bleicher:1999xi} results, indicated by the hashed red lines, are above the experimental measurements. For energies higher than 20 GeV the UrQMD results overshoot the HRG baseline as well. In the right panel of Fig.~\ref{fig_STAR} the $\kappa_{4}/\kappa_{2}$ results are presented. For central collisions at higher energies the experimental results approach the HRG limit of unity, while at around 27 GeV an evident dip emerges, rendering the non-monotonic energy dependence. The possible further increase of the $\kappa_{4}/\kappa_{2}$ ratio, in the context of probing the proximity of the critical point, should be accompanied with this behaviour. Unfortunately the apparent increase of the measured $\kappa_{4}/\kappa_{2}$ value at $\sqrt{s_{\mathrm {NN}}} = 7.7$~GeV is not statistically significant. 

It should be mentioned that, even if the dramatic increase of $\kappa_{2}/\kappa_{2}$ at lower energies would be in place, it would still be not enough to claim the discovery of the critical point. In order to allow firm conclusions, further measurements at even lower energies would be necessary. With the upgraded STAR detector the approved BES-II program aims at significantly improved statistics (about factor of 20), increased acceptance in both $p_{T}$, $y$ and extended  measurements down to $\sqrt{s_{\mathrm {NN}}} = 3$~GeV in the fixed-target configuration.

Recently the STAR collaboration reported on their first measurements of the $\kappa{6}/\kappa{2}$ ratio~\cite{Pandav:2020uzx}. At $\sqrt{s_{\mathrm {NN}}} = 200$~GeV the measured $\kappa{6}/\kappa{2}$ values are negative, while at $\sqrt{s_{\mathrm {NN}}} = 54$~GeV they remain positive. This sign change between the two energies is at odds wit the latest LQCD calculations~\cite{Bazavov:2020bjn, Bollweg:2020yum, Ding:2020rtq}. The experimental measurements of 5$^{th}$ order cumulants would shed light on these discrepancies.

\subsection{Results from NA61/SHINE}
The NA61/SHINE experiment aims at exploring the phase structure of strongly interacting matter by performing 2-dimensional scan in beam momentum (13-150/158A GeV/c) and sizes of the colliding systems. In the left and right panels of Fig.~\ref{fig_NA61} the measured $\kappa_{3}/\kappa_{2}$ and $\kappa_{4}/\kappa_{2}$ ratios of  net-charge distribution are presented, respectively. The consistency of both ratios with the corresponding EPOS results, shown by solid black lines, indicates that the measurements are essentially driven by conservation laws. However, the corresponding cumulant ratios for separate charges cannot be described with the EPOS model anymore~\cite{Mackowiak-Pawlowska:2020glz}. This disagreement indicates that further systematic analysis is needed to fully understand the system size dependence  of fluctuation measurements performed by the NA61/SHINE collaboration. 

\subsection{Results from HADES}
As mentioned in the previous section, it is essential to extend the fluctuation measurements towards lower energies. Recently the HADES experiment performed systematic studies of proton multiplicity distribution in A--Au collisions at $\sqrt{s_{\mathrm {NN}}} = 2.4$~GeV. Together with the cumulnats of proton distribution the HADES collaboration investigated multi-particle correlators. The integrated correlators are directly linked to cumulants of distributions~\cite{Bzdak:2016sxg}. In particular, in order to understand  higher order cumulants, it is essential to separately investigate differential properties of correlators. Indeed, as argued in~\cite{Ling:2015yau}, the functional behaviour of  cumulants on selected rapidity range $\Delta y$ allows to distinguish between long- and short-range correlations. At HADES energies, however, it is more appropriate to use mean number of protons instead of $\Delta y$. In this representation short- and long-rage correlations lead to $C_{n}\propto \left<N_{p}\right>$ and $C_{n}\propto \left<N_{p}\right>^{n}$ scalings, respectively.
In Fig.~\ref{fig_HADES} the efficiency and volume corrected integrated multi-particle  correlators are presented as measured by the HADES collaboration~\cite{Adamczewski-Musch:2020slf}. The data are fit with power-law functions $C_{n}(\left<N_{p}\right>)=C_{0}\left<N_{p}\right>^{\alpha}$, where the exponent $\alpha$ and normalization constant $C_{0}$ are fit parameters. For the most central events (0-5\%) the obtained values of $\alpha$ parameter are found  to be close to the order of the corresponding multi-particle correlator, i.e, $\alpha \thickapprox  n$ for all $C_{n}$, with $n$= 2, 3, 4. This indicates that long range correlations dominate the cumulants measured in Au--Au collisions by the HADES collaboration.

\section{Summary}
In summary, several measurements performed by the ALICE, HADES, NA61/SHINE and STAR collaborations are discussed. The correlations persistent in normalized second cumulant R$_1$, as measured by the ALICE collaboration, are induced by collisions in the very early phase of the Pb--Pb interaction. After accounting for baryon number conservation, the ALICE data are in agreement with the corresponding second cumulants of the Skellam distribution, consistent with the LQCD calculations at a pseudo-critical temperature of about 156 MeV~\cite{Bazavov:2020bjn}. The power-law behaviour of multi-particle correlators, reported by the HADES collaboration, suggest long range rapidity correlations. The STAR data on $\kappa_{4}/\kappa_{2}$ exhibits non-monotonic behaviour at around $\sqrt{s_{\mathrm {NN}}} = 20$~GeV. However, the sign change of $\kappa_{6}/\kappa_{2}$, as reported by the STAR collaboration, is not consistent with the latest LQCD calculations.  The measurements of net-charge fluctuations in small systems performed by the  NA61/SHINE collaboration are in good agreement with the corresponding EPOS results.  Near future challenges will be precision measurements of higher moments  at RHIC, LHC, SIS as well as at facilities such as FAIR at GSI and NICA at JINR  and their connection to fundamental QCD properties.

\section*{Acknowledgments}
This work is part of and supported by the DFG Collaborative Research Centre "SFB 1225 (ISOQUANT)". 





\bibliographystyle{elsarticle-num}
\bibliography{references}

\begin{thebibliography}{10}
\expandafter\ifx\csname url\endcsname\relax
  \def\url#1{\texttt{#1}}\fi
\expandafter\ifx\csname urlprefix\endcsname\relax\def\urlprefix{URL }\fi
\expandafter\ifx\csname href\endcsname\relax
  \def\href#1#2{#2} \def\path#1{#1}\fi

\bibitem{Koch:2008ia}
V.~Koch, {Hadronic Fluctuations and Correlations}, 2010, pp. 626--652.
\newblock \href {http://arxiv.org/abs/0810.2520} {\path{arXiv:0810.2520}},
  \href {http://dx.doi.org/10.1007/978-3-642-01539-7\_20}
  {\path{doi:10.1007/978-3-642-01539-7\_20}}.

\bibitem{Adamczyk:2013dal}
L.~Adamczyk, et~al., {Energy Dependence of Moments of Net-proton Multiplicity
  Distributions at RHIC}, Phys.\ Rev.\ Lett. 112 (2014) 032302.
\newblock \href {http://arxiv.org/abs/1309.5681} {\path{arXiv:1309.5681}},
  \href {http://dx.doi.org/10.1103/PhysRevLett.112.032302}
  {\path{doi:10.1103/PhysRevLett.112.032302}}.

\bibitem{Bazavov:2020bjn}
A.~Bazavov, et~al., {Skewness, kurtosis and the 5th and 6th order cumulants of
  net baryon-number distributions from lattice QCD confront high-statistics
  STAR data}, Phys.\ Rev.\ D 101~(7) (2020) 074502.
\newblock \href {http://arxiv.org/abs/2001.08530} {\path{arXiv:2001.08530}},
  \href {http://dx.doi.org/10.1103/PhysRevD.101.074502}
  {\path{doi:10.1103/PhysRevD.101.074502}}.

\bibitem{Stephanov:2004wx}
M.~A. Stephanov, {QCD phase diagram and the critical point}, Prog.\ Theor.\
  Phys.\ Suppl. 153 (2004) 139--156.
\newblock \href {http://arxiv.org/abs/hep-ph/0402115}
  {\path{arXiv:hep-ph/0402115}}, \href
  {http://dx.doi.org/10.1142/S0217751X05027965}
  {\path{doi:10.1142/S0217751X05027965}}.

\bibitem{LQCD1}
A.~Bazavov, et~al., {The chiral and deconfinement aspects of the QCD
  transition}, Phys.\ Rev.\ D 85 (2012) 054503.
\newblock \href {http://arxiv.org/abs/1111.1710} {\path{arXiv:1111.1710}},
  \href {http://dx.doi.org/10.1103/PhysRevD.85.054503}
  {\path{doi:10.1103/PhysRevD.85.054503}}.

\bibitem{LQCD2}
S.~Borsanyi, Z.~Fodor, J.~N. Guenther, S.~K. Katz, K.~K. Szabo, A.~Pasztor,
  I.~Portillo, C.~Ratti, {Higher order fluctuations and correlations of
  conserved charges from lattice QCD}, JHEP 10 (2018) 205.
\newblock \href {http://arxiv.org/abs/1805.04445} {\path{arXiv:1805.04445}},
  \href {http://dx.doi.org/10.1007/JHEP10(2018)205}
  {\path{doi:10.1007/JHEP10(2018)205}}.

\bibitem{Friman:2011pf}
B.~Friman, F.~Karsch, K.~Redlich, V.~Skokov, {Fluctuations as probe of the QCD
  phase transition and freeze-out in heavy ion collisions at LHC and RHIC},
  Eur.\ Phys.\ J.\ C 71 (2011) 1694.
\newblock \href {http://arxiv.org/abs/1103.3511} {\path{arXiv:1103.3511}},
  \href {http://dx.doi.org/10.1140/epjc/s10052-011-1694-2}
  {\path{doi:10.1140/epjc/s10052-011-1694-2}}.

\bibitem{Andronic:2017pug}
A.~Andronic, P.~Braun-Munzinger, K.~Redlich, J.~Stachel, {Decoding the phase
  structure of QCD via particle production at high energy}, Nature 561~(7723)
  (2018) 321--330.
\newblock \href {http://arxiv.org/abs/1710.09425} {\path{arXiv:1710.09425}},
  \href {http://dx.doi.org/10.1038/s41586-018-0491-6}
  {\path{doi:10.1038/s41586-018-0491-6}}.

\bibitem{Andronic:2018qqt}
A.~Andronic, P.~Braun-Munzinger, B.~Friman, P.~M. Lo, K.~Redlich, J.~Stachel,
  {The thermal proton yield anomaly in Pb-Pb collisions at the LHC and its
  resolution}, Phys.\ Lett.\ B 792 (2019) 304--309.
\newblock \href {http://arxiv.org/abs/1808.03102} {\path{arXiv:1808.03102}},
  \href {http://dx.doi.org/10.1016/j.physletb.2019.03.052}
  {\path{doi:10.1016/j.physletb.2019.03.052}}.

\bibitem{Adamczyk:2017iwn}
L.~Adamczyk, et~al., {Bulk Properties of the Medium Produced in Relativistic
  Heavy-Ion Collisions from the Beam Energy Scan Program}, Phys.\ Rev.\ C
  96~(4) (2017) 044904.
\newblock \href {http://arxiv.org/abs/1701.07065} {\path{arXiv:1701.07065}},
  \href {http://dx.doi.org/10.1103/PhysRevC.96.044904}
  {\path{doi:10.1103/PhysRevC.96.044904}}.

\bibitem{Braun-Munzinger:2016yjz}
P.~Braun-Munzinger, A.~Rustamov, J.~Stachel, {Bridging the gap between
  event-by-event fluctuation measurements and theory predictions in
  relativistic nuclear collisions}, Nucl.\ Phys.\ A 960 (2017) 114--130.
\newblock \href {http://arxiv.org/abs/1612.00702} {\path{arXiv:1612.00702}},
  \href {http://dx.doi.org/10.1016/j.nuclphysa.2017.01.011}
  {\path{doi:10.1016/j.nuclphysa.2017.01.011}}.

\bibitem{Skokov:2012ds}
V.~Skokov, B.~Friman, K.~Redlich, {Volume Fluctuations and Higher Order
  Cumulants of the Net Baryon Number}, Phys.\ Rev.\ C 88 (2013) 034911.
\newblock \href {http://arxiv.org/abs/1205.4756} {\path{arXiv:1205.4756}},
  \href {http://dx.doi.org/10.1103/PhysRevC.88.034911}
  {\path{doi:10.1103/PhysRevC.88.034911}}.

\bibitem{Acharya:2019izy}
S.~Acharya, et~al., {Global baryon number conservation encoded in net-proton
  fluctuations measured in Pb-Pb collisions at $\sqrt{s_{\rm NN}}$ = 2.76 TeV.
  }\href {http://arxiv.org/abs/1910.14396} {\path{arXiv:1910.14396}}.

\bibitem{Braun-Munzinger:2019yxj}
P.~Braun-Munzinger, A.~Rustamov, J.~Stachel, {The role of the local
  conservation laws in fluctuations of conserved charges. }\href
  {http://arxiv.org/abs/1907.03032} {\path{arXiv:1907.03032}}.

\bibitem{Adamczewski-Musch:2020slf}
J.~Adamczewski-Musch, et~al., {Proton number fluctuations in $\sqrt{s_{NN}}$ =
  2.4 GeV Au+Au collisions studied with HADES. }\href
  {http://arxiv.org/abs/2002.08701} {\path{arXiv:2002.08701}}.

\bibitem{Rustamov:2017lio}
A.~Rustamov, {Net-baryon fluctuations measured with ALICE at the CERN LHC},
  Nucl.\ Phys.\ A 967 (2017) 453--456.
\newblock \href {http://arxiv.org/abs/1704.05329} {\path{arXiv:1704.05329}},
  \href {http://dx.doi.org/10.1016/j.nuclphysa.2017.05.111}
  {\path{doi:10.1016/j.nuclphysa.2017.05.111}}.

\bibitem{Gazdzicki:2011xz}
M.~Gazdzicki, K.~Grebieszkow, M.~Mackowiak, S.~Mrowczynski, {Identity method to
  study chemical fluctuations in relativistic heavy-ion collisions}, Phys.\
  Rev.\ C 83 (2011) 054907.
\newblock \href {http://arxiv.org/abs/1103.2887} {\path{arXiv:1103.2887}},
  \href {http://dx.doi.org/10.1103/PhysRevC.83.054907}
  {\path{doi:10.1103/PhysRevC.83.054907}}.

\bibitem{Gorenstein:2011hr}
M.~Gorenstein, {Identity Method for Particle Number Fluctuations and
  Correlations}, Phys.\ Rev.\ C 84 (2011) 024902, [Erratum: Phys.Rev.C 97,
  029903 (2018)].
\newblock \href {http://arxiv.org/abs/1106.4473} {\path{arXiv:1106.4473}},
  \href {http://dx.doi.org/10.1103/PhysRevC.84.024902}
  {\path{doi:10.1103/PhysRevC.84.024902}}.

\bibitem{Rustamov:2012bx}
A.~Rustamov, M.~Gorenstein, {Identity Method for Moments of Multiplicity
  Distribution}, Phys.\ Rev.\ C 86 (2012) 044906.
\newblock \href {http://arxiv.org/abs/1204.6632} {\path{arXiv:1204.6632}},
  \href {http://dx.doi.org/10.1103/PhysRevC.86.044906}
  {\path{doi:10.1103/PhysRevC.86.044906}}.

\bibitem{Arslandok:2018pcu}
M.~Arslandok, A.~Rustamov, {TIdentity module for the reconstruction of the
  moments of multiplicity distributions}, Nucl.\ Instrum.\ Meth.\ A 946 (2019)
  162622.
\newblock \href {http://arxiv.org/abs/1807.06370} {\path{arXiv:1807.06370}},
  \href {http://dx.doi.org/10.1016/j.nima.2019.162622}
  {\path{doi:10.1016/j.nima.2019.162622}}.

\bibitem{Bzdak:2012an}
A.~Bzdak, V.~Koch, V.~Skokov, {Baryon number conservation and the cumulants of
  the net proton distribution}, Phys.\ Rev.\ C 87~(1) (2013) 014901.
\newblock \href {http://arxiv.org/abs/1203.4529} {\path{arXiv:1203.4529}},
  \href {http://dx.doi.org/10.1103/PhysRevC.87.014901}
  {\path{doi:10.1103/PhysRevC.87.014901}}.

\bibitem{Gyulassy:1994ew}
M.~Gyulassy, X.-N. Wang, {HIJING 1.0: A Monte Carlo program for parton and
  particle production in high-energy hadronic and nuclear collisions}, Comput.\
  Phys.\ Commun. 83 (1994) 307.
\newblock \href {http://arxiv.org/abs/nucl-th/9502021}
  {\path{arXiv:nucl-th/9502021}}, \href
  {http://dx.doi.org/10.1016/0010-4655(94)90057-4}
  {\path{doi:10.1016/0010-4655(94)90057-4}}.

\bibitem{Dumitru:2008wn}
A.~Dumitru, F.~Gelis, L.~McLerran, R.~Venugopalan, {Glasma flux tubes and the
  near side ridge phenomenon at RHIC}, Nucl.\ Phys.\ A 810 (2008) 91--108.
\newblock \href {http://arxiv.org/abs/0804.3858} {\path{arXiv:0804.3858}},
  \href {http://dx.doi.org/10.1016/j.nuclphysa.2008.06.012}
  {\path{doi:10.1016/j.nuclphysa.2008.06.012}}.

\bibitem{Aamodt:2011mr}
K.~Aamodt, et~al., {Two-pion Bose-Einstein correlations in central Pb-Pb
  collisions at $\sqrt{{s}_{NN}} =$ 2.76 TeV}, Phys.\ Lett.\ B 696 (2011)
  328--337.
\newblock \href {http://arxiv.org/abs/1012.4035} {\path{arXiv:1012.4035}},
  \href {http://dx.doi.org/10.1016/j.physletb.2010.12.053}
  {\path{doi:10.1016/j.physletb.2010.12.053}}.

\bibitem{Arslandok:2020mda}
M.~Arslandok, {Recent results on net-baryon fluctuations in ALICE}, in: {28th
  International Conference on Ultrarelativistic Nucleus-Nucleus Collisions},
  2020.
\newblock \href {http://arxiv.org/abs/2002.03906} {\path{arXiv:2002.03906}}.

\bibitem{Almasi:2017bhq}
G.~A. Almasi, B.~Friman, K.~Redlich, {Baryon number fluctuations in chiral
  effective models and their phenomenological implications}, Phys.\ Rev.\ D
  96~(1) (2017) 014027.
\newblock \href {http://arxiv.org/abs/1703.05947} {\path{arXiv:1703.05947}},
  \href {http://dx.doi.org/10.1103/PhysRevD.96.014027}
  {\path{doi:10.1103/PhysRevD.96.014027}}.

\bibitem{Adam:2020unf}
J.~Adam, et~al., {Net-proton number fluctuations and the Quantum Chromodynamics
  critical point}\href {http://arxiv.org/abs/2001.02852}
  {\path{arXiv:2001.02852}}.

\bibitem{Braun-Munzinger:2018yru}
P.~Braun-Munzinger, A.~Rustamov, J.~Stachel, {Experimental results on
  fluctuations of conserved charges confronted with predictions from canonical
  thermodynamics}, Nucl.\ Phys.\ A 982 (2019) 307--310.
\newblock \href {http://arxiv.org/abs/1807.08927} {\path{arXiv:1807.08927}},
  \href {http://dx.doi.org/10.1016/j.nuclphysa.2018.09.074}
  {\path{doi:10.1016/j.nuclphysa.2018.09.074}}.

\bibitem{Bleicher:1999xi}
M.~Bleicher, et~al., {Relativistic hadron hadron collisions in the
  ultrarelativistic quantum molecular dynamics model}, J.\ Phys.\ G 25 (1999)
  1859--1896.
\newblock \href {http://arxiv.org/abs/hep-ph/9909407}
  {\path{arXiv:hep-ph/9909407}}, \href
  {http://dx.doi.org/10.1088/0954-3899/25/9/308}
  {\path{doi:10.1088/0954-3899/25/9/308}}.

\bibitem{Pandav:2020uzx}
A.~Pandav, {Measurement of cumulants of conserved charge multiplicity
  distributions in Au+Au collisions from the STAR experiment. }\href
  {http://arxiv.org/abs/2003.12503} {\path{arXiv:2003.12503}}.

\bibitem{Bollweg:2020yum}
D.~Bollweg, F.~Karsch, S.~Mukherjee, C.~Schmidt, {Higher order cumulants of net
  baryon-number distributions at non-zero $\mu_B$}, in: {28th International
  Conference on Ultrarelativistic Nucleus-Nucleus Collisions}, 2020.
\newblock \href {http://arxiv.org/abs/2002.01837} {\path{arXiv:2002.01837}}.

\bibitem{Ding:2020rtq}
H.-T. Ding, {New developments in lattice QCD on equilibrium physics and phase
  diagram}, in: {28th International Conference on Ultrarelativistic
  Nucleus-Nucleus Collisions}, 2020.
\newblock \href {http://arxiv.org/abs/2002.11957} {\path{arXiv:2002.11957}}.

\bibitem{Mackowiak-Pawlowska:2020glz}
M.~Mackowiak-Pawlowska, {NA61/SHINE results on fluctuations and correlations at
  CERN SPS energies}, in: {28th International Conference on Ultrarelativistic
  Nucleus-Nucleus Collisions}, 2020.
\newblock \href {http://arxiv.org/abs/2002.04847} {\path{arXiv:2002.04847}}.

\bibitem{Bzdak:2016sxg}
A.~Bzdak, V.~Koch, N.~Strodthoff, {Cumulants and correlation functions versus
  the QCD phase diagram}, Phys.\ Rev.\ C 95~(5) (2017) 054906.
\newblock \href {http://arxiv.org/abs/1607.07375} {\path{arXiv:1607.07375}},
  \href {http://dx.doi.org/10.1103/PhysRevC.95.054906}
  {\path{doi:10.1103/PhysRevC.95.054906}}.

\bibitem{Ling:2015yau}
B.~Ling, M.~A. Stephanov, {Acceptance dependence of fluctuation measures near
  the QCD critical point}, Phys.\ Rev.\ C 93~(3) (2016) 034915.
\newblock \href {http://arxiv.org/abs/1512.09125} {\path{arXiv:1512.09125}},
  \href {http://dx.doi.org/10.1103/PhysRevC.93.034915}
  {\path{doi:10.1103/PhysRevC.93.034915}}.

\end{thebibliography}







\end{document}